\documentclass[11pt]{article}
\usepackage{RR}
\RRNo{6266}
\usepackage{latexsym}
\usepackage{graphicx}
\usepackage{theorem}
\usepackage{algorithm}
\usepackage[noend]{algorithmic}
\newcommand{\remove}[1]{}
\newfloat{Algorithm}{hpt}{lop}

\newtheorem{lemma}{Lemma}
\newtheorem{theorem}{Theorem}
\newenvironment{proof}{\textbf{Proof:}}{\hfill$\Box$}
\newtheorem{definition}{Definition}

\RRdate{July 2007}
\RRtitle{Sur l'Auto-stabilisation de Robots Mobiles dans un Graphe}
\RRetitle{On the Self-stabilization of Mobile Robots in Graphs}
\titlehead{Stabilizing Robots in Graphs}
\RRauthor{L\'{e}lia Blin$^{\dagger}$ \and Maria Gradinariu Potop-Butucaru$^{^\star}$ \and S\'{e}bastien Tixeuil$^{\ddag}$\\
\vspace{.5cm}
\small{
$^\dagger$ Universit\'{e} d'Evry, IBISC-CNRS, France

$^\star$ Universit\'{e} Pierre et Marie Curie-Paris 6, LIP6-CNRS, INRIA REGAL, France

$^{\ddag}$ Universit\'{e} Paris Sud-XI, LRI-CNRS, INRIA Grand Large, France
}
}

\authorhead{L. Blin, M. Gradinariu Potop-Butucaru, S. Tixeuil}

\RRresume{
L'auto-stabilisation est une technique g\'{e}n\'{e}rique pour tol\'{e}rer toute d\'{e}faillance transitoire dans un syst\`{e}me distribu\'{e}. Les robots (ou agents) mobiles constituent l'un des mod\`{e}les \'{e}mergents de l'informatique distribu\'{e}e du fait de leur ressemblance avec les entit\'{e}s biologiques autonomes.
La contribution de cet article est triple. D'abord, nous pr\'{e}sentons un nouveau mod\`{e}le pour l'\'{e}tude des entit\'{e}s mobiles dans des r\'{e}seaux sujets \`{a} des d\'{e}faillances transitoires. Notre mod\`{e}le diff\`{e}re du mod\`{e}le classique des robots car les robots ont des contraintes sur les chemins qu'ils peuvent emprunter, et du mod\`{e}le classique des agents mobiles car le nombre d'agents reste fixe pendant toute la dur\'{e}e de l'ex\'{e}cution du protocole. Ensuite, dans ce mod\`{e}le, nous \'{e}tudions la possibilit\'{e} de l'existence d'algorithmes auto-stabilisants quand ces algorithmes sont ex\'{e}cut\'{e}s par des robots mobiles \'{e}voluant dans un graphe. Nous nous concentrons sur les briques de bases des syst\`{e}mes \`{a} base d'agents et de robots~: le nomage et l'\'{e}lection d'un chef. Conform\'{e}ment \`{a} l'intuition, quand aucune contrainte n'est faite sur le r\'{e}seau et les param\`{e}tres de l'ex\'{e}cution, les deux probl\`{e}mes sont impossibles \`{a} r\'{e}soudre. Enfin, quand les hypoth\`{e}ses minimales pour r\'{e}soudre ces deux probl\`{e}mes sont disponibles, nous proposons des solutions d\'{e}terministes et probabilistes pour les deux probl\`{e}mes, et montrons que ces deux probl\`{e}mes sont \'{e}quivalents au moyen d'une r\'{e}duction algorithmique.
}

\RRabstract{
Self-stabilization is a versatile technique to withstand any transient
fault in a distributed system. Mobile robots (or agents) 
are one of the emerging trends in distributed computing as they mimic autonomous biologic entities. 
The contribution of this paper is threefold. First, we present a new
model for studying mobile entities in networks subject 
to transient faults. Our model differs from the classical robot model
because robots have constraints about the paths they 
are allowed to follow, and from the classical agent model because the
number of agents remains fixed throughout the execution 
of the protocol. Second, in this model, we study the possibility of
designing self-stabilizing algorithms when those 
algorithms are run by mobile robots (or agents) evolving on a
graph. We concentrate on the core building blocks of 
robot and agents problems: naming and leader election. Not
surprisingly, when no constraints are given on the network 
graph topology and local execution model, both problems are impossible to solve. 
Finally, using minimal hypothesis with respect to impossibility
results, we provide deterministic and 
probabilistic solutions to both problems, and show equivalence of
these problems by an algorithmic reduction mechanism.
}

\RRmotcle{agents mobiles, robots mobiles, graphes, auto-stabilisation, \'{e}lection
d'un chef, nomage}
\RRkeyword{mobile agents, mobile robots, graphs, self-stabilization,
leader election, naming}
\RRprojet{Grand Large}
\RRtheme{\THNum}
\URFuturs

\begin{document}
\makeRR

\section{Introduction}

A large panel of recent research in Distributed Computing focused on solving problems using mobile entities, often denoted by the term of \emph{robots} or \emph{agents}. Those entities typically evolve in the network (that comprises a fixed set of nodes forming a particular topology) to provide services, either to the user of the network or to its core components. With the advent of large-scale networks that involve a total number of components in the order of the million, two particular issues were stressed: \emph{(i)} the resources used by the agents should be kept to a minimum given a particular problem (see \emph{e.g.}~\cite{FIRT06bc}), and \emph{(ii)} the fault and attack tolerance capabilities are of premium importance. Most of the works on fault and attack tolerance with mobile agents deals with \emph{external} threats, \emph{i.e.} the faulty part of the system or the attacker is not an agent itself. For example, several papers (\emph{e.g.}~\cite{DFPS02c}) investigate the black hole search problem, where mobile entities must cooperate to find a hostile node of the network that destroys every mobile entity traversing it. In an orthogonal manner, decontamination and graph searching papers (\emph{e.g.}~\cite{FFN05c}) consider the chasing of hostile mobile entities that are harmful to the nodes but not to the agents.

In this paper, we consider the novel problem of dealing with faults and attacks that hit the mobile entities themselves, that is, the threat is \emph{internal}. More precisely, we consider that an arbitrary transient fault or attack hits the system (both nodes and mobile entities), and devise algorithmic solutions to recover from those faults and attacks. The faults and attacks are \emph{transient} in the sense that there exists a point from which they don't appear any more. In practice, it is sufficient that the faults and attacks are sporadic enough for the network to provide useful services most of the time. In this context, \emph{self-stabilization}~\cite{D00b} is an elegant approach to forward recovery from transient faults and attacks as well as initializing a large-scale system. Informally, a self-stabilizing systems is able to recover from any transient fault or attack in finite time, without restricting the nature or the span of those faults and attacks.

\paragraph{Related works}

Mobile (software) agents on graphs were studied in the context of self-stabilization \emph{e.g.} 
in \cite{G00c,HM01c,BHS01c,DSW02c}, but the implicit model is completely different from ours. In the aforementioned 
works, agents are software entities that are exchanged through messages between processes (that are located in the nodes of the network), 
and thus can be destroyed, duplicated, and created at will. In \cite{G00c,BHS01c}, a single agent is assumed at a given time, 
and this agent is responsible for stabilizing a simultaneously running (classical \emph{aka} non-stabilizing) distributed algorithm. 
In \cite{HM01c}, exactly $n$ agents are supposed to traverse a $n$ sized tree network infinitely often, by means of a \emph{swap} 
primitive that swaps agents located at two neighboring nodes. In \cite{DSW02c}, the authors consider dynamic evolving networks and 
rely on random walks to ensure proper agent traversal; again, the purpose of the agent protocol is to ensure that a single agent 
stabilizes the system. 
By contrast, in this paper, we focus on the self-stabilization of the agents themselves, and our model 
keeps the number of agents fixed for the whole life of the network. 
In a similar way, in \cite{DGR05} is studied the resource allocation
problem in a model where agents (the resource consumers) are mobile and the
resources form a fixed network. In this model agents join and leave the resource network at
will. However, contrary to the curent work the model proposed in
\cite{DGR05} assumes that agents and nodes have unique identifiers.

Self-stabilizing mobile (hardware) robots in 2-dimensional space were recently studied in \emph{e.g.} \cite{suzuki96distributed,prencipe01corda,flocchini00distributed}. 
Here a fixed set of $k$ mobile robots are able to move unconstrained, yet are not able to communicate other than by 
viewing the relative position of other robots. 
The presented algorithms are oblivious in the sense that between any two activations of a particular robot, 
the previous state of the robot is cleared. As such, those algorithms are inherently 
self-stabilizing, since any scheduling for execution will reset the state of a robot.
In this model several problems have been studied  under different assumptions on the environment 
(schedulers, fault-tolerance, robots visibility, accuracy of compasses): circle formation, pattern 
formation, gathering, leader election, scattering.
However, the lack of digital communication between the robots somewhat limits the kind of problems that can be 
solved and a broad class of impossibility results have been obtained \cite{suzuki96distributed, 
prencipe01corda,flocchini00distributed,DGMP06, AP04}. 
In this paper, we introduce non-oblivious robots (agents) in the context of self-stabilization and enable digital 
communication between robots (either because they are located at the same node, or by using so-called whiteboards) 
to solve more elaborate tasks (\emph{e.g.} naming and leader election), yet we restrict the motion capabilities of the robots (only edges of a given graph can be followed).

A third related model in the area of self-stabilization is that of Population Protocols (see \emph{e.g.} \cite{AngluinADFP2006,AngluinAFJ2005,DBLP:conf/podc/AngluinADFP04}).
In this model, finite-state agents interact in pairs chosen by an adversary, with
both agents updating their state according to a joint transition function. 
For each such transition function,
the resulting population protocol is said to stably compute a predicate on the initial states of the agents if,
after sufficiently many interactions in a fair execution, all agents converge to having the correct value of the
predicate. It was proved that this model permits to compute problems that can be expressed through Presburger arithmetic. Our model permits as well to express joint transition functions between agents located at the same node, but also (indirectly) between agents that are hosted by the same node at different moments through the whiteboard abstraction.

\paragraph{Our contribution}

The contribution of this paper is threefold. First, we present a new
model for studying mobile entities in networks subject to transient
fault. Our model differs from the classical robot model because robots
have constraints about the paths they are allowed to follow, and from
the classical agent model because the number of agents remains fixed
throughout the execution of the protocol. Second, in this model, we
study the possibility of designing self-stabilizing algorithms when
those algorithms are run by mobile robots (or agents) evolving on a
graph. We concentrate on the core building blocks of robot and agents
problems: naming and leader election. Not surprisingly, when no
constraints are given on the network graph topology and local
execution model, both problems are impossible to solve. Finally, using
minimal hypotheses with respect to impossibility results, we provide
deterministic and probabilistic solutions to both problems, and show
equivalence of these problems by an algorithmic reduction
mechanism. From a theoretical perspective, our results complement the
widely known possible \emph{vs.} impossible problems in anonymous
distributed systems (see \emph{e.g.}~\cite{YK96ja,YK96jb}). From a
practical perspective, our symmetry breaking algorithm enables to
solve other problems such as 
gathering (see \cite{DFKP06j}) that have known solutions 
when mobile entities have unique identifiers. 

\paragraph{Outline}
In Section~\ref{sec:model}, we present the computing model that is introduced in this paper. Section~\ref{sec:impossible} provides impossibility results that justify the assumptions made in subsequent sections. Section~\ref{sec:dnaming} presents a deterministic algorithm for naming in acyclic networks with half-duplex links, while Section~\ref{sec:pnaming} provides a probabilistic naming algorithm for general networks. Section~\ref{sec:leader} shows that the naming and leader election problem are equivalent (by reduction of one problem to the other). Concluding remarks are presented in Section~\ref{sec:conclusion}.

\section{Model}
\label{sec:model}

The network is modeled as a connected graph $G=(V,E)$, where $V$ is a set of nodes, and $E$ is a set of edges. We assume that nodes have local distinct labels for links, but no assumption is made about the labeling process. Nodes also maintain a so-called \emph{whiteboard} which can store a fixed amount of information. We distinguish between \emph{acyclic} networks (\emph{i.e.} trees) and \emph{cyclic} networks (\emph{i.e.} that contain at least one cycle).

\emph{Agents} (or \emph{robots}) are entities that move between neighboring nodes in the network. A link is \emph{full-duplex} if two agents located at neighboring nodes can exchange their position at the same time, crossing the same link in opposite directions. A link is \emph{half-duplex} if only one direction can be used at a given time.
We assume that $k$ agents are present in the network at any time. Also, each agent is modeled by an automata whose state space is sufficient to hold a identifier that is unique in the network (\emph{i.e.} the state space is at least $k$ states). An agent may move from one node to one of the node's neighbors based on the following information: \emph{(i)} the current state of the agent, \emph{(ii)} the current state of other agents located at the same node, \emph{(iii)} the local link labels of the current node (and possibly the label of the incoming link used by the agent to reach the node), and \emph{(iv)} the memory stored at the node (the whiteboard).

A \emph{configuration} of the system is the product of all agents locations and states and all whiteboards contents. 
The behavior of the system is essentially \emph{asynchronous}, in the sense that there is no bound on the number of moves that an agent can make between any two moves of another agent. There is one notable exception: when two (or more) 
agents are at the same node, the execution order is decided by the host node.
In the following we assume that no two agents located at the same node execute their actions concurrently.
However, agents located at different nodes may execute their actions concurrently. 
So, in any configuration of the system, the scheduler may chose any subset of nodes that hold at least 
one agent: then, in one atomic step all chosen nodes execute the code of all their host agents. The pseudo-code of each node (unless otherwise stated) is formally presented as Algorithm~\ref{alg:move_node}.

\begin{Algorithm}
\begin{center}
\begin{verbatim}
foreach agent on node
  agent.execute()
end foreach
\end{verbatim}
\end{center}
\caption{Pseudo-code at node $i$}
\label{alg:move_node}
\end{Algorithm}

We assume that the scheduler is \emph{fair} in the sense that if a node holds at least one agent, it is eventually scheduled for execution by the scheduler. A \emph{round} starting from configuration $c$ is the minimum time until all nodes that hold at least one agent in $c$ have been activated by the scheduler.
We now define the naming and leader election problems that we focus on in this paper: 

\begin{definition}[Naming]
\label{def:naming}
Let $S$ be a system with $k$ agents. The system $S$ satisfies the {\it naming} specification if all $k$ agents eventually have an unique identifier (no two agents in $S$ share the same identifier).
\end{definition}

In the leader election problem agents have either the leader role or the follower role.

\begin{definition}[Leader Election]
\label{def:le}
Let $S$ be a system with $k$ agents. The system $S$ satisfies the {\it leader election} 
specification if an unique agent eventually has the leader role 
and all the agents agents have the follower role.
\end{definition}

Our goal is to withstand transient failures. For this purpose, we assume that every ``changing'' aspect of the network can be arbitrarily modified in the initial configuration of the system. These ``changing'' aspects include: \emph{(i)} the agent states, \emph{(ii)} the agent positions, \emph{(iii)} the whiteboard states.

\begin{definition}[Self-stabilizing Problem]
Let $S$ be a system with $k$ agents. The system $S$ satisfies the {\it naming} or the {\it leader election} 
specification in a self-stabilizing way 
if Definition \ref{def:naming} respectively Definition \ref{def:le} are verified 
in spite of an arbitrary initial configuration.
\end{definition}

\section{Impossibility results}
\label{sec:impossible}

The results in this section are negative. We consider networks where agents have infinite memory, and nodes have whiteboards with infinite memory. Also, the scheduling is constrained in the sense that at every step, all nodes that hold at least one agent are scheduled for execution. With our assumptions, the initial memory of every agent is supposed to be identical, and the initial content of each whiteboard is supposed to be identical as well. Agents are anonymous and deterministic. 

\begin{theorem}
\label{theo:impossibility}
Deterministic naming or leader election of mobile agents is impossible in cyclic networks, even assuming synchronous scheduling, and infinite memory for each agent and whiteboard.
\end{theorem}

\begin{proof}
Assume the topology of the network is a cycle. The proof idea follows the lines of impossibility results found in~\cite{A80c}. Intuitively, assume a cyclic network in a symmetric initial configuration with two agents. Assume the agents have both the same identifier (or leader) variable
and all the whiteboards in the network are initialized with the same arbitrary values. Since agents execute the same deterministic algorithm, there exists an execution of the system refereed in the following as $e$ such that all configurations appearing in this execution are symmetrical with respect to the agents view. Since a configuration solving the naming problem is asymmetrical with respect to the agents view, and the fact that $e$ does not contain asymmetrical configurations, there exists an execution of the system that never reaches a naming or leader election configuration. 

In the following we construct a symmetric execution $e$ with respect to the agents view. Without restraining the generality assume only two agents in the network placed such that agents have exactly the same view and the same initial state $s_0$.
Since the two agents have the same view, start in identical states, and execute the same deterministic algorithm, they execute the same action. So, both agents will reach exactly the same state $s_0$ (in case the agents do not change their state) or $s_1 \neq s_0$. 
In the new state the agents have the same view of the network and the same state so they execute again the same action. Either, the two agents keep the same state or they both change to a new state $s_2$. The agents can repeat this game infinitely often.
Overall, there is an infinite execution where the two agents pass exactly through the same states in the same time $(s_0)^{*}(s_1)^{*}(s_2)^* \ldots$ and the system never reaches an asymmetrical configuration. 
\end{proof}

\begin{theorem}
\label{thm:full-duplex}
Deterministic naming or leader election of mobile agents is impossible in acyclic networks with full-duplex links, even assuming synchronous scheduling, and infinite memory for each agent and whiteboard.
\end{theorem}

\begin{proof}
Consider a network consisting of two nodes $u$ and $v$ linked by edge $e$. Assume that there is an agent at each node. Initially all agents are in the same state, and all nodes whiteboards are in the same state. Also, the local labeling of edges is the same at each node. So, the network is completely symmetric, from an agent point of view. Now, each time an agent is scheduled for execution, it may update its own state, update the whiteboard, or (inclusive) move to the other node. Now assume that the scheduling of the agents is synchronous, this means that at every step, the state of each agent remains identical, the state of each whiteboard remains identical, and the relative position of each agent with respect to the view of the other agent 
remains the same. Overall, symmetry can not be broken by a deterministic agent algorithm, and naming or leader election can not be achieved.
\end{proof}

\section{Self-stabilizing deterministic naming in acyclic networks with half-duplex links}
\label{sec:dnaming}

In the following we propose a deterministic self-stabilizing algorithm for agents naming in acyclic networks with half-duplex links. In networks with $k$ agents the algorithm uses $O(\log(k))$ bits memory per agent and $O(k\log(\Delta k))$ per node, where $\Delta$ is the maximum degree of the network. 

Each agent has a state that includes a software identifier $\mathtt{id}$ (``software'' meaning that this identifier can be corrupted), that is represented by some integer. Each node has a whiteboard (that can be corrupted as well) that can store up to $k$ $2$-tuples $\langle \mathtt{id},\mathtt{edge} \rangle$. The $\mathtt{id}$ part of the $2$-tuple denotes the integer identifier of an agent, and the $\mathtt{edge}$ part of the $2$-tuple denotes a local edge identifier. This whiteboard is meant to represent the identifiers of the latest $k$ agents with distinct identifiers that visited the node, along with the corresponding outgoing edge they took last time they visited the node. Each node provides to the agents that visit the node some helper functions to access the whiteboard:

\begin{itemize}
\item $\mathtt{edge}(i)$ returns the edge that is associated to $i$ if $i$ is present in the whiteboard, and $\mathit{nil}$ if $i$ is not in the whiteboard.
\item $\mathtt{visit}(i,j)$ sets the edge $j$ to be associated to agent $i$ if $i$ is in the whiteboard, or adds the entry $(i,j)$ to the whiteboard if $i$ is not present. If the whiteboard already contains $k$ tuples with distinct identifiers, the least recently updated one is dropped from the whiteboard. After updating the whiteboard, the agent exits the node through port $j$.
\item $\mathtt{new}$ returns a new identifier that does not exists in the whiteboard.
\end{itemize}

When arriving at a node, an agent checks whether the node has its identifier in its whiteboard. If it is not present, it adds its identifier and erases one of the identifiers if there are more than $k$ identifiers on the node's whiteboard (assuming FIFO order). If it is present, then either there is another agent with the same identifier at the current node, or the agent is the only one with its identifier. In the first case, the first agent to be executed by the node simply picks up a new identifier, and initiates a Eulerian traversal. In the second case, the agent checks if the last outgoing edge followed by an agent with its identifier is the same as the current incoming edge. In the case it points to another edge (which means there is a discrepancy, whatever its cause), the node simply follows the path corresponding to its identifier, trying to confront the other agent with the same identifier (if such agent exists). Finally, when an agent enters a node through the expected edge, it continues performing a Eulerian tour of the tree, \emph{e.g.} by choosing the outgoing edge that is next in the Eulerian tour, \emph{i.e.} edge $(j+1)~\mathrm{mod}~\delta$, if $j$ was the incoming edge.
Formally, the algorithm that is executed by each agent is presented as Algorithm~\ref{alg:agent_tree}. 

\begin{Algorithm}
\begin{center}
\begin{verbatim}
id: integer
execute() {
  if ( edge(id) == nil ) 
    visit( id, 0 ) // add self, will exit through port 0 by default
  else if ( exists agent j on node i such that j.id == id )    
    id = new // Not present or somebody else has the same id 
    visit( id, 0 ) // add self, will exit through port 0 by default
  else if ( edge(id) != incoming edge ) )
    visit( id, edge( id ) ) // Follow possible conflicting agent
  else 
    visit( id, edge( id ) + 1 mod delta ) // continue Eulerian traversal
}
\end{verbatim}
\end{center}
\caption{Deterministic agent code for tree networks}
\label{alg:agent_tree}
\end{Algorithm}

We prove self-stabilization by defining a predicate for \emph{legitimate} configurations, then proving \emph{(i)} every computation starting from a legitimate configuration is correct (see Appendix), and \emph{(ii)} every computation starting from an arbitrary configuration eventually reaches a legitimate configuration. 
 
\begin{definition}[Legitimate configuration]
A legitimate configuration for Algorithm \ref{alg:agent_tree} satisfies the three following properties: \emph{(i)} all agents have distinct ``software'' identifiers, \emph{(ii)} all whiteboards tuples contain only actual agent identifiers, and \emph{(iii)} all whiteboards tuples are consistent with actual agent locations.
\end{definition}

In the following we prove that starting from any configuration, the system converges to a legitimate configuration.

\begin{lemma}
\label{lemma:infinitvisit}
An agent with identifier $i$ eventually visits every node infinitely often.
\end{lemma}

\begin{proof}
Assume the contrary, \emph{i.e.} there exists a time in the execution
 where at least one node $u$ never gets visited by any agent with identifier $i$. In turn, this means that for every
 neighbor $v$ of $u$, either $v$ is never visited by an agent with identifier $i$ (and the argument can be repeated on 
the neighbors of $v$), or $v$ is visited infinitely often but the agent never takes the edge toward $u$ (shortened as $e_u$). 
The only way for an agent $a$ with identifier $i$ not to take $e_u$ is \emph{(i)} to exit through edge number $0$ 
(and that edge is not $e_u$), \emph{(ii)} to follow the path of a (supposedly) other agent with identifier $i$ that did 
not take $e_u$, or \emph{(iii)} to never take $e_u$ by performing a Eulerian traversal of the graph (\emph{i.e.} 
the agent never arrives by port $(e_u - 1 ~\mathrm{mod} ~\delta)$). Cases \emph{(i)} and \emph{(ii)} can be executed 
only a finite number of times (since there are $k$ agents), so this implies that the node is visited infinitely often
 by agents that properly execute the $(\mathtt{edge}(i)+1 ~\mathrm{mod} ~\delta)$ rule and never exit through $e_u$. Since the network is acyclic, this last case is impossible.
\end{proof}

\begin{lemma}
\label{lemma:circuit}
If two agents have the same identifier they eventually meet within $O(m)$ rounds, 
where $m$ is the number of edges of the graph.
\end{lemma}

\begin{proof}
Note that a node that keeps 
its identifier either follows the apparent path of a supposed other node, or performs a Eulerian traversal of the tree. 
Intuitively, the proof goes as follows.
The apparent path may be either fake (it leads to a node that does not have identifier $i$ in its whiteboard, or that 
does not contain another agent with identifier $i$) or real (the path leads to an agent whose identifier is $i$). 
If the path is fake, it is nevertheless finite, and the agent will perform only a finite number of steps to reach the 
end of the path and realize it is fake. When an agent realizes a path is fake, it executes the Eulerian traversal algorithm. 
When a node switches to the Eulerian traversal algorithm, its path becomes real. Now, if the path is real, the agent chases 
a real agent in an acyclic network, and the path information is correct. Since the network is acyclic and the links are 
half duplex, the two agents are bound to meet each other.

An agent follows a fake path when the information on whiteboards erroneously indicates the presence of another agent 
with the same identifier. In order to check and correct the information in the whiteboards an agents, in the worst case, has 
to visit every node in network. In order to perform the traversal, each edge is visited at most twice.
Hence the complexity of the traversal is $O(m)$.
\end{proof}

Note that after an agent visited each node of the graph at least once, all whiteboards 
are coherent with the agent direction and identifier.

\begin{lemma}[Convergence]
Starting from any arbitrary initial configuration with $k$ agents, any computation eventually 
reaches a legitimate configuration in $O(k m)$ rounds.
\end{lemma}

\begin{proof}
First, we observe that no identifier that is initially present in the network on some agent is ever removed from the network. 
This is due to the fact that an agent only changes its identifier when observing that another agent at the same node has the exact 
same identifier. Since agents execute their code sequentially (activated by nodes), the first activated agent with a conflicting 
identifier change its identifier, and the other agent remain unchanged (unless there are more 
that two agents at the same node with the same identifier).

Now, we prove that starting from any initial configuration, the number of distinct identifiers only increases until it reaches $k$. 
Initially the number of distinct identifiers is at least $1$. Suppose that there exists some integer $j$ ($1\leq j < k$) such that 
there exists $j$ distinct identifiers in the network. Now, after finite time, $O(j m)$, all $j$ identifiers are present in each whiteboard in the network 
(see Lemma \ref{lemma:circuit}). Since $j<k$, there exist at least two agents with the same identifier. By the above argument, 
two agents with the same identifier are 
to meet within finite time, $O(m)$. When this is done, one of the agents will change its identifier to a new identifier. Since all $j$ 
identifiers are in the whiteboards and are regularly refreshed, a new identifier (not in the existing $j$ set) will be solicited 
by the agent, and the number of total identifiers in the network rises to $j+1$. 
By induction hypothesis, the number of distinct identifier eventually reaches $k$ after $O(k m)$ rounds.

When all agents have distinct identifiers, they all traverse all the network infinitely often 
(see Lemma \ref{lemma:infinitvisit}). When each of 
them has traversed the network at least once, the paths to the agents are all correct with respect to their current position. 
As a consequence, the configuration is legitimate.
\end{proof}

\begin{lemma}[Correctness]
Every computation of Algorithm~\ref{alg:agent_tree} that starts from a legitimate configuration satisfies the Naming problem specification.
\end{lemma}

\begin{proof}
Since all $k$ agents have distinct identifiers, and all $k$ identifiers are present in all whiteboards, the whiteboards do not contain any spurious identifier information. So, an agent arriving at a node always finds its own identifier, with proper incoming edge. As a result, agents never change identifier, whiteboards never drop existing identifiers, and edge information is kept accurate, so that every agent performs Eulerian traversal of the tree forever.
\end{proof}

\section{Probabilistic naming in arbitrary networks}
\label{sec:pnaming}

In this section we assume an weaker model where agents cannot
communicate via whiteboard and the network is arbitrary with full-duplex links. 
Theorems~\ref{theo:impossibility} and \ref{thm:full-duplex} provide impossibility results related to 
deterministic naming in this model. In the following
we show the possibility of probabilistic naming.
The idea is to make every agent randomly move in the network. Anytime two agents that are located at 
the same node have the same identifier, each agent randomly chooses a new identifier. If there are 
several agents at the same node with distinct identifiers, the random walk is continued.

Each node provides to the agents that visit the node some helper functions:

\begin{itemize}
\item $\mathtt{random}(S)$ returns a random element from set $S$.
\item $\mathtt{visit}(j)$ makes the agent exit the node through port $j$.
\end{itemize}

\begin{Algorithm}
\begin{center}
\begin{verbatim}
id: integer
execute() { 
   if (there exists agent a such that a.id = id at node i)
       id := random( 1..k )
   else
       visit( random( 0..degree(i) )
}
\end{verbatim}
\end{center}
\caption{Probabilistic agent code executed at node $i$ for arbitrary networks}
\label{alg:probabilistic_agent}
\end{Algorithm}

Algorithm \ref{alg:probabilistic_agent} presents the core of our algorithm for probabilistic 
agent naming. In the presentation, a random value for the identifier is assumed to be between $1$ and $k$, but an upper bound on $k$ may 
also be used to boost stabilization time (\emph{e.g.} $k^2$). The
proof of correctness can be found in the appendix.

\begin{definition}[Legitimate configuration]
A configuration is legitimate if and only if all agents have distinct identifiers.
\end{definition}

\begin{lemma}[Convergence]
Starting from an arbitrary initial configuration, the network eventually reaches a legitimate configuration.
The expected stabilization time is $O(kn^3)$ where $k$ is the number of agents in the network.
\end{lemma}

\begin{proof}
We first make the two following observations:
\begin{enumerate}
\item when two agents with two different identifiers meet at the same 
node in the network, their random walk is unaffected by the meeting;
\item when two agents with the same identifier meet at the same node in the network, 
they stop moving until at least one of them randomly picked a new identifier.
\end{enumerate}
In an arbitrary initial configuration, a pair of agents $(u,v)$ in the network may share the same identifier. 
We consider occurrences of meetings of two agents or more at the same node in the network. When random walks are unbiased, 
the meeting time between any two agents is $O(n^3)$, \cite{TW91}. Here we do not consider the meeting time between agents with 
different identifiers. Instead we consider the first occurrence of a meeting involving two or more agents of the same color. 
When this occurs, the two agents draw a random coin and get a random identifier. 
With probability at least $\frac{1}{k}$, the drawn fresh random identifier is unique in the whole network. 
So, anytime two agents with the same identifier meet, there is a positive probability 
that they both get a unique identifier in the system. Anytime this 
happens, the number of agents who share their color with at least one other agent decreases by one. As a result, with probability 
one, a configuration where all agents have unique identifiers is reached, and remains thereafter.
The stabilization time $O(kn^3)$.
\end{proof}

\begin{lemma}[Correctness]
Every computation of Algorithm~\ref{alg:probabilistic_agent} that starts from a legitimate configuration satisfies the naming problem specification.
\end{lemma}

\begin{proof}
Assume all identifiers are distinct for all agents, then the ``if'' clause is never falsified, 
so the identifier of the agent is never changed. As a result, the configuration remains legitimate.
\end{proof}

\section{Naming and leader election}
\label{sec:leader}
In this section, we consider the relationship between the aforementioned naming problem, and the 
leader election problem, where the network must eventually 
reach a configuration where exactly one agent is elected and all others are non-elected. 

\subsection{From naming to leader election}

We first observe that given a naming of $k$ robots in the network, it is easy to come up with a leader election protocol. 
\begin{enumerate}
\item In our deterministic protocol, whiteboards are used to register all identifiers used in the system by the agents. When an agent arrives at some node, it checks from 
the whiteboard if its identifier is maximum in the whiteboard, and becomes elected if so. If its identifier is not maximum in the node's whiteboard, the agent becomes non-elected. 
After stabilization of the naming algorithm, all whiteboards contain the exact identifiers used in the network, which means that all whiteboards contain 
the same identifiers in the system. So, if an agent has maximal identifier on one whiteboard, 
it has maximum identifier on all whiteboards. This guarantees the correctness of the leader election protocol.
\item In our probabilistic protocol, eventually all nodes have distinct identifiers in the network. If the exact value of $k$ was used in the algorithm, then a 
node can simply checks its identifier against $k$ to detect if it is the leader or not. If only an upper bound on $k$ was used, then a more complicated process 
is required. The procedure is as follows: each node stores the identifiers of the last $k-1$ distinct agents it last saw; then if its identifier is maximum 
among those identifiers, it becomes elected, and remains non-elected otherwise. The leader status is updated anytime the list of the $k-1$ distinct encounters is 
modified. After stabilization of the naming algorithm, all identifiers are distinct, so a non-biased random walk is performed by each agent. Then each agents meets 
every other agent regularly within polynomial time. Overall, after polynomial time, every agent has met every other agent and stored their identifiers in its local 
memory. When every agent has all other agent identifiers in its local memory, the leader status remain correct and unchanged. The memory cost 
of the algorithm is $O(k\log(k))$ per agent and the time complexity is polynomial.

An alternative to this algorithm is as follows. Each agent performs a random walk in the network (at each node the agent chooses with equal probability (1/node degree) the 
next edge to visit).
In \cite{Feige95} it is proved that the expected time for a random walk to cover all nodes of a 
graph is $O(n \log(n))$. Each time an agent visits a node it marks in the node table its identifier 
if it is not present. If the agent identifier is the maximum in the table then the agent is the leader otherwise 
it keeps the follower status. The memory complexity of the algorithm 
is $O(k \log(k))$ per node and the expected time complexity is $O(k n \log(n))$.
\end{enumerate}

\subsection{From leader election to naming}

Now consider the reverse problem of solving the naming problem given a leader in the group of agents. Our solution is presented as 
Algorithms~\ref{alg:agent_leader} and \ref{alg:node_leader}.
For simplicity, we assume that the leader agent is identified by a 
special symbol that is not in the domain on non-leader agents identifiers and which can be recognized by the node as the leader mark. 
First we assume that each node, when activating agents, gives lower priority to the leader agent (\emph{i.e.} 
the code of the leader agent, if present on the node, is executed last). The intuition of the algorithm is as follows: the leader agent simply performs a Eulerian 
traversal of the tree, and is not influenced by the other nodes. On its way during each traversal, the leader leaves in the \texttt{edge} variable of each 
traversed whiteboard the outgoing edge it used to exit last time it visited the node. In a legitimate situation, those \texttt{edge} variables constitute a 
tree pointing toward the current location of the leader. The rationale for the non-leader nodes is as follows: \emph{(i)} when the leader is not present on 
the same node, the non-leader node simply follows the \texttt{edge} left by the leader, trying to reach it, and \emph{(ii)} when the leader is present on the 
same node, the non-leader agent first checks against duplicate identifiers of non-leader agents located at the same leader-based node, and pick up a new fresh 
identifier if needed, then they take the same outgoing edge as the leader, in order to always remain at the same node as the leader agent. Since the network 
is acyclic and the links are half-duplex, every non-leader node eventually meets the leader, and once met, they never leave the leader. So, eventually, 
the leader agent collects all non-leader agents at its current location. When all agents are co-located at the same node 
and check that no duplicate identifiers exist, the naming process is finished.

\begin{Algorithm}
\begin{center}
\begin{verbatim}
id: integer
execute() {
  if ( leader ) 
    edge := edge + 1 mod delta // follow the Eulerian traversal
    visit( edge )
  else
    if ( leader is present on the same node ) 
      if (id is conflicting among present agents on the node)
        id := new // take fresh identifier
      visit( edge + 1 mod delta ) // take same exit as that of the leader
    else
      visit( edge ) // follow leader
}
\end{verbatim}
\end{center}
\caption{Deterministic agent code for naming in tree networks}
\label{alg:agent_leader}
\end{Algorithm}

\begin{Algorithm}
\begin{center}
\begin{verbatim}
foreach non-leader agent on node
  agent.execute()
end foreach
leader.execute()
\end{verbatim}
\end{center}
\caption{Deterministic node code for naming in tree networks}
\label{alg:node_leader}
\end{Algorithm}

\begin{definition}[Legitimate configuration]
A configuration of the network is legitimate if it satisfies the following properties: \emph{(i)} all non-leader agents have distinct identifiers, \emph{(ii)} all agents are located in the same node, and \emph{(iii)} all \texttt{edge} whiteboards point toward the node that contain all agents.
\end{definition}

\begin{lemma}[Correctness]
Starting from a legitimate configuration, the naming problem is solved.
\end{lemma}

\begin{proof}
In a legitimate configuration, all agents have distinct identifiers. From the code of the algorithm, an agent may 
change its identifier only when discovering that it shares the same identifier with another agent. 
As a result, an agent never changes its identifier onwards, and the naming problem is solved.
\end{proof}

\begin{lemma}[Convergence]
Starting from an arbitrary initial configuration, a legitimate configuration is eventually reached.
\end{lemma}

\begin{proof}
We first prove that eventually, all \texttt{edge} whiteboards point toward the node that contain the leader. We observe that the leader behavior 
does not depend on the behavior of the non-leader agents. Second, when the leader leaves a node (whatever the initialization of the whiteboard of 
this node may be), the \texttt{edge} whiteboard of this node will always point toward the leader agent onwards (the network is acyclic, so the 
leader agent may only come back though this edge), and the next time the \texttt{edge} whiteboard is modified, it will advertise the current last 
taken edge by the leader agent. Our second observation is that whatever the initialization of the whiteboards, the leader agent always perform a 
Eulerian traversal of the network. As a result, all nodes are eventually visited by the leader node, and when all nodes have been visited at least 
once, all \texttt{edge} whiteboards are pointing toward the leader agent.

The second step of the proof is to show that any non-leader agent eventually meets the leader agent. Since the \texttt{edge} whiteboards all point 
toward the leader agent, and that non-leader agents that are not located on the same node as the leader simply follow the \texttt{edge} whiteboards, 
they always move toward the leader agent. Since the network is acyclic and the links are half-duplex, the leader agent and any non-leader agent are 
bound to meet within $O(m)$ rounds. Now, when a non-leader agent meets the leader agent, their moving behavior remains the same hereafter 
(\emph{i.e.} the leader and the non-leader agents follow exactly the same path at the same moment--when the node they are both located on is activated). 
So, eventually, within $O(k m)$ rounds all agents are located at the same node at a given moment, 
and remain located at the same node hereafter (though the node they are located
 changes anytime it is scheduled for execution). 

When all agents are gathered at the same node, non-leader nodes (that are executed in sequence when the node is activated) simply choose different identifiers. 
After one such node activation, all agents have distinct identifiers, are gathered at the same node, and all \texttt{edge} whiteboards are pointing to them. 
As a consequence, the configuration is legitimate.
\end{proof}

\section{Concluding remarks}
\label{sec:conclusion}

In this paper, we introduced the problem of self-stabilizing mobile robots in graphs, and presented deterministic and probabilistic solutions to the 
problems of naming, and leader election among robots. From a practical point of view, the main difference between the two solutions is that the 
deterministic solution uses a whiteboard (\emph{i.e.} a local memory available at every node that the agents can use to communicate with others) 
while the probabilistic one does not. An interesting open question
that is raised by this work is the trade-off between 
whiteboard availability and randomness capabilities. In addition, it would be of theoretical interest to prove that the computational power of our model is strictly greater (in terms of predicates that can be computed) than the Population Protocol model.

\bibliographystyle{plain}
\bibliography{naming}

\begin{thebibliography}{10}

\bibitem{AP04}
N.~Agmon and D.~Peleg.
\newblock Fault-tolerant gathering algorithms for autonomous mobile robots.
\newblock In {\em Proc. 15th Annual ACM-SIAM Symposium on Discrete Algorithms
  (SODA 2004)}, pages 1070--1078, New Orleans, LA, USA, January 2004.

\bibitem{A80c}
Dana Angluin.
\newblock Local and global properties in networks of processors (extended
  abstract).
\newblock In {\em STOC}, pages 82--93. ACM, 1980.

\bibitem{DBLP:conf/podc/AngluinADFP04}
Dana Angluin, James Aspnes, Zo{\"e} Diamadi, Michael~J. Fischer, and Ren{\'e}
  Peralta.
\newblock Computation in networks of passively mobile finite-state sensors.
\newblock In {\em PODC}, pages 290--299, 2004.

\bibitem{AngluinADFP2006}
Dana Angluin, James Aspnes, Zo{\"e} Diamadi, Michael~J. Fischer, and Ren\'e
  Peralta.
\newblock Computation in networks of passively mobile finite-state sensors.
\newblock {\em Distributed Computing}, pages 235--253, March 2006.

\bibitem{AngluinAFJ2005}
Dana Angluin, James Aspnes, Michael~J. Fischer, and Hong Jiang.
\newblock Self-stabilizing population protocols.
\newblock In {\em Principles of Distributed Systems; 9th International
  Conference, OPODIS 2005; Pisa, Italy; December 2005; Revised Selected
  Papers}, volume 3974 of {\em Lecture Notes in Computer Science}, pages
  103--117, December 2005.

\bibitem{BHS01c}
J.~Beauquier, T.~Herault, and E.~Schiller.
\newblock {Easy Stabilization with an Agent}.
\newblock {\em 5th Workshop on Self-Stabilizing Systems (WSS)}, 2194:35--51.

\bibitem{DGR05}
Ajoy~Kumar Datta, Maria Gradinariu, and Michel Raynal.
\newblock Stabilizing mobile philosophers.
\newblock {\em Inf. Process. Lett.}, 95(1):299--306, 2005.

\bibitem{DGMP06}
Xavier D{\'e}fago, Maria Gradinariu, St{\'e}phane Messika, and Philippe~Raipin
  Parv{\'e}dy.
\newblock Fault-tolerant and self-stabilizing mobile robots gathering.
\newblock In {\em DISC}, pages 46--60, 2006.

\bibitem{DFKP06j}
Anders Dessmark, Pierre Fraigniaud, Dariusz~R. Kowalski, and Andrzej Pelc.
\newblock Deterministic rendezvous in graphs.
\newblock {\em Algorithmica}, 46(1):69--96, 2006.

\bibitem{DFPS02c}
Stefan Dobrev, Paola Flocchini, Giuseppe Prencipe, and Nicola Santoro.
\newblock Searching for a black hole in arbitrary networks: optimal mobile
  agent protocols.
\newblock In {\em PODC}, pages 153--161, 2002.

\bibitem{D00b}
S.~Dolev.
\newblock {\em Self-stabilization}.
\newblock MIT Press, March 2000.

\bibitem{DSW02c}
S.~Dolev, E.~Schiller, and J.~Welch.
\newblock {Random walk for self-stabilizing group communication in ad-hoc
  networks}.
\newblock {\em Reliable Distributed Systems, 2002. Proceedings. 21st IEEE
  Symposium on}, pages 70--79, 2002.

\bibitem{Feige95}
Uriel Feige.
\newblock A tight upper bound on the cover time for random walks on graphs.
\newblock {\em Random Struct. Algorithms}, 6(1):51--54, 1995.

\bibitem{flocchini00distributed}
P.~Flocchini, G.~Prencipe, N.~Santoro, and P.~Widmayer.
\newblock Distributed coordination of a set of autonomous mobile robots.
\newblock {\em IVS, pages 480-485, 2000.}, 2000.

\bibitem{FFN05c}
Fedor~V. Fomin, Pierre Fraigniaud, and Nicolas Nisse.
\newblock Nondeterministic graph searching: From pathwidth to treewidth.
\newblock In Joanna Jedrzejowicz and Andrzej Szepietowski, editors, {\em MFCS},
  volume 3618 of {\em Lecture Notes in Computer Science}, pages 364--375.
  Springer, 2005.

\bibitem{FIRT06bc}
Pierre Fraigniaud, David Ilcinkas, Sergio Rajsbaum, and S{\'e}bastien Tixeuil.
\newblock The reduced automata technique for graph exploration space lower
  bounds.
\newblock In Oded Goldreich, Arnold~L. Rosenberg, and Alan~L. Selman, editors,
  {\em Essays in Memory of Shimon Even}, volume 3895 of {\em Lecture Notes in
  Computer Science}, pages 1--26. Springer, 2006.

\bibitem{G00c}
S.~Ghosh.
\newblock {Agents, distributed algorithms, and stabilization}.
\newblock {\em Computing and Combinatorics (COCOON 2000), Springer LNCS}, pages
  242--251, 2000.

\bibitem{HM01c}
T.~Herman and T.~Masuzawa.
\newblock {Self-Stabilizing Agent Traversal}.
\newblock {\em WSS01 Proceedings of the Fifth International Workshop on
  Self-Stabilizing Systems, Springer LNCS}, 2194:152--166, 2001.

\bibitem{prencipe01corda}
G.~Prencipe.
\newblock Corda: Distributed coordination of a set of autonomous mobile robots.
\newblock {\em Proc. ERSADS, pages 185--190, May 2001.}, 2001.

\bibitem{suzuki96distributed}
I.~Suzuki and M.~Yamashita.
\newblock Distributed anonymous mobile robots---formation and agreement
  problems.
\newblock {\em Proceedings of the 3rd International Colloquium on Structural
  Information and Communication Complexity (SIROCCO '96), Siena, Italy, June
  1996.}, 1996.

\bibitem{TW91}
Prasad Tetali and Peter Winkler.
\newblock On a random walk problem arising in self-stabilizing token
  management.
\newblock In {\em PODC}, pages 273--280, 1991.

\bibitem{YK96ja}
Masafumi Yamashita and Tsunehiko Kameda.
\newblock Computing on anonymous networks: Part i-characterizing the solvable
  cases.
\newblock {\em IEEE Trans. Parallel Distrib. Syst.}, 7(1):69--89, 1996.

\bibitem{YK96jb}
Masafumi Yamashita and Tsunehiko Kameda.
\newblock Computing on anonymous networks: Part ii-decision and membership
  problems.
\newblock {\em IEEE Trans. Parallel Distrib. Syst.}, 7(1):90--96, 1996.

\end{thebibliography}

\end{document}